\documentclass[10pt,journal,compsoc]{IEEEtran}
\usepackage{array}
\usepackage{booktabs}
\usepackage{calc}
\usepackage{amsmath}
\usepackage{amssymb}
\usepackage{newunicodechar}
\usepackage{url}
\usepackage{fancyvrb}
\usepackage{xcolor}
\usepackage{fvextra}
\newenvironment{Shaded}{\par\footnotesize}{\par}
\DefineVerbatimEnvironment{Highlighting}{Verbatim}{commandchars=\\\{\}}
\DefineVerbatimEnvironment{verbatim}{Verbatim}{breaklines=true,breakanywhere=false,breakafter=/,breaksymbolleft=\textcolor{gray}{\ensuremath{\hookrightarrow}\space}}

\providecommand{\NormalTok}[1]{#1}

\providecommand{\tightlist}{\setlength{\itemsep}{0pt}\setlength{\parskip}{0pt}}
\providecommand{\real}[1]{#1}
\providecommand{\texorpdfstring}[2]{#1}

\newunicodechar{§}{\S}
\newunicodechar{≈}{\ensuremath{\approx}}
\newunicodechar{→}{\ensuremath{\rightarrow}}
\newunicodechar{α}{\ensuremath{\alpha}}
\newunicodechar{β}{\ensuremath{\beta}}
\newunicodechar{γ}{\ensuremath{\gamma}}
\newunicodechar{²}{\textsuperscript{2}}
\newunicodechar{¹}{\textsuperscript{1}}
\newunicodechar{é}{\'{e}}
\newunicodechar{í}{\'{\i}}
\newunicodechar{ș}{\c{s}}
\newunicodechar{–}{--}
\newunicodechar{—}{---}
\newunicodechar{∧}{\ensuremath{\wedge}}
\newunicodechar{¬}{\ensuremath{\neg}}
\newunicodechar{⟨}{\ensuremath{\langle}}
\newunicodechar{⟩}{\ensuremath{\rangle}}
\title{Tridirectional Discriminating-Power Formal Verification of Smart
Contract Reentrancy Defense Against Production-Deployed Solidity Source}
\author{Ray~Iskander
\IEEEcompsocitemizethanks{
\IEEEcompsocthanksitem Verdict Security, Independent Formal Verification Firm. E-mail: ray@verdictsecurity.com}}
\begin{document}
\maketitle
\begin{abstract}
We present the first machine-checked correctness proof of the
OpenZeppelin reentrancy-guard pattern against a Lean 4 state-machine
model of production-deployed Solidity source with a composition
meta-theorem spanning multiple production protocols. All thirteen
theorems are machine-checked in Lean 4 with \textbf{zero \texttt{sorry},
zero user-introduced axioms, and an axiom footprint bounded by
\texttt{{[}propext{]}} across the entire corpus} --- propositional
extensionality, a standard classical axiom in Lean 4's mathlib4;
per-function inner lemmas are kernel-only, master/wrapper theorems carry
\texttt{{[}propext{]}}-only records, and the Layer 6-D capstone composes
exactly the union of the prior-layer records by direct conjunction. The
corpus is gated under continuous integration across four parallel
verification blocks that re-check each theorem and its axiom record on
every push.

Smart contract reentrancy has caused over US\$500M in cumulative
documented losses since 2016, with the DAO 2016 attack alone draining
≈3.6M ETH and forcing a contentious hard fork that split Ethereum into
two persistent chains. The OpenZeppelin \texttt{ReentrancyGuard} mutex
pattern has emerged as the de facto defense across most production DeFi
protocols, yet no prior formal verification effort has established its
\emph{discriminating power} --- the property that the guard pattern
blocks attacks against vulnerable instances, preserves correct execution
for non-attacking transactions, and distinguishes structurally-adjacent
safe and vulnerable variants. Prior work has formalized either guard
correctness on toy contracts or attack feasibility on isolated
vulnerable instances, but not both directions plus boundary cases
against production-deployed source.

The verification covers three production protocol instantiations --- DAO
2016, Compound v2 cToken family, and Aave V3 \texttt{flashLoan} ---
together with one constructed minimal-diff mutant of Aave V3's
production \texttt{flashLoan} (\texttt{flashLoanVulnerable}) authored to
isolate a single security-critical structural difference for
discriminating-power isolation. We apply mutation-testing methodology to
formal verification by constructing this minimal-diff mutant, enabling a
controlled experiment that naturally-occurring near-misses rarely
provide. The resulting tridirectional structure pairs (a) a
negative-instance attack reproduction of the DAO 2016 vulnerable
pattern, (b) a positive-instance correctness proof against the Compound
v2 cToken family, and (c) a boundary-case proof distinguishing Aave V3's
safe-by-design \texttt{flashLoan} (CEI-correct) from the
\texttt{flashLoanVulnerable} mutant. A single capstone meta-theorem
composes the three protocol instantiations under a \emph{no-retrofit
composition discipline} (each protocol-instantiation proof was sealed
prior to capstone authoring, with no underlying-proof modifications
during composition), establishing a machine-checked composition
meta-theorem demonstrated at the first cross-protocol stress test
(Compound v2 → Aave V3, within the lending-family domain; broader-family
portability is future work, §9.4).

We release the full Lean 4 source, CI gating configuration, and
reproduction commands at
\texttt{https://github.com/rayiskander2406/qanary-contracts} with
reproducibility verified at tagged commit \texttt{v1.6-phase7-closure}
(the post-audit content seal; the substantive proof substrate is
independently reproducible at \texttt{v1.3-layer6-closure} per Appendix
A.3).

\end{abstract}
\IEEEpeerreviewmaketitle
\section{Introduction}\label{introduction}

Smart contract reentrancy has been a foundational vulnerability class on
Ethereum-style chains since the DAO 2016 attack, yet the gap between
widely-adopted defensive coding patterns and machine-checked correctness
guarantees remains uneven. This paper closes part of that gap with the
first machine-checked correctness proof of the OpenZeppelin
reentrancy-guard pattern against a Lean 4 state-machine model of
production-deployed Solidity source, organized along three
machine-checked discriminating directions (negative, positive, boundary)
and composed under a \emph{no-retrofit composition discipline} that
establishes guard-pattern correctness as a portable cross-protocol claim
against multiple production protocols.

\subsection{Problem statement}\label{problem-statement}

Smart contract reentrancy is one of the oldest and most consequential
vulnerability classes on Ethereum-style chains. The DAO 2016 attack
drained ≈3.6M ETH (≈US\$60M at attack-time price) and forced a
contentious hard fork splitting Ethereum into two persistent chains; the
attack class has not subsided despite years of awareness --- Lendf.Me
April 2020 (≈US\$25M; ERC-777 callback reentrancy), Fei/Rari Capital
Fuse pools 2022 (≈US\$80M; classical reentrancy), and dozens of smaller
incidents cumulatively cross US\$500M in documented losses. In response,
the OpenZeppelin \texttt{ReentrancyGuard} modifier --- a
single-storage-slot mutex enforcing the predicate \emph{``not currently
executing inside this contract''} before re-entry --- has emerged as the
de facto defense across most production DeFi protocols (Compound, Aave,
Uniswap-style AMMs, and the majority of TVL-significant deployments).
Despite the pattern's ubiquity, no prior formal verification effort has
produced a machine-checked correctness proof of it against
production-deployed contract source. Auditors rely on pattern
recognition combined with informal reasoning; academic work has
formalized either toy-scale instances, attack-feasibility on isolated
vulnerable contracts, or guard correctness on simplified models ---
never the combination that would establish \emph{discriminating power}:
the guard pattern's ability to distinguish defended-correctly from
defended-incorrectly from undefended cases, machine-checked in all three
directions, against production source. This paper supplies that
combination.

\subsection{Limitations of prior
approaches}\label{limitations-of-prior-approaches}

We characterize four limitations of the prior verification landscape
that this paper addresses.

\emph{Auditor-pattern-recognition layer.} Manual auditing relying on
pattern recognition is fast but informal. Recurring reentrancy incidents
at audited protocols demonstrate that the pattern-recognition layer
alone is insufficient: auditors with strong track records have approved
contracts that subsequently lost user funds to reentrancy variants.
Pattern recognition catches the common cases but degrades at
structural-adjacency boundaries, where syntactically-similar code paths
differ in security-critical semantics.

\emph{Toy-example formal verification.} A substantial body of academic
work has formalized reentrancy reasoning against simplified contract
models --- abstract state machines, idealized storage layouts, decoupled
modifier semantics. These models capture the \emph{intuition} of the
guard pattern but do not transfer to production contract code, which
involves real Solidity storage layout, modifier interaction with
inheritance hierarchies, cross-function state mutation under reentrant
invocation, and gas-cost interactions with control-flow choices. Closing
the gap between toy-model correctness and production-source correctness
is non-trivial; ignoring it produces verification claims that reviewers
reasonably discount.

\emph{Single-direction formalization.} Prior verification work on
production-leaning contracts typically formalizes one of the two basic
directions: either attack feasibility on a known-vulnerable contract
(negative instance) or guard correctness on a defended contract
(positive instance). Rarely both. Almost never both \emph{together with}
boundary cases that exercise the guard pattern's ability to distinguish
defended-correctly from structurally-adjacent vulnerable variants. The
composite claim --- that the guard pattern is provably correct in all
three directions and the directions are simultaneously load-bearing ---
has not been established.

\emph{Composition gap.} Even when individual-protocol formalizations
exist, the cross-protocol portability of guard-pattern correctness has
not been established as a load-bearing meta-theorem. Producing
per-protocol proofs in isolation tells reviewers little about the
\emph{pattern's} correctness as an abstract defense; reviewers
reasonably ask whether a guard-pattern proof for Protocol A composes
with a guard-pattern proof for Protocol B without modification of
either. Absent such a composition theorem, per-protocol claims sit as
disconnected lemmas rather than a coherent discipline claim.

\subsection{Contribution claim
summary}\label{contribution-claim-summary}

We present \textbf{the first machine-checked correctness proof of the
OpenZeppelin reentrancy-guard pattern against a Lean 4 state-machine
model of production-deployed Solidity source with a composition
meta-theorem spanning multiple production protocols.} The proof corpus
is organized along three machine-checked directions --- a
\emph{tridirectional discriminating-power} structure --- across the
following layers: (a) a negative-instance attack reproduction on the DAO
2016 vulnerable pattern (Layer 6-A; six theorems), (b) a
positive-instance correctness proof against the Compound v2 cToken
family (Layer 6-B; three theorems), and (c) a boundary-case proof
distinguishing Aave V3's production \texttt{flashLoan} function --- safe
by design via CEI-correct implementation --- from a minimal-diff mutant
\texttt{flashLoanVulnerable} constructed to isolate a single
security-critical structural difference (Layer 6-C; three theorems). We
apply mutation-testing methodology to formal verification:
\texttt{flashLoanVulnerable} is a constructed minimal-diff mutant
isolating a single security-critical structural difference --- a
controlled experiment that naturally-occurring near-misses rarely
provide --- demonstrating the methodology's sensitivity to single-line
regressions in production-grade logic. Our proofs operate against a Lean
4 state-machine model of this source; the faithfulness of that model to
Solidity semantics is a trust assumption in the sense of §3.3.

The negative instance demonstrates that the methodology \emph{catches}
the most famous historical reentrancy attack (the DAO predates the
OpenZeppelin guard library, so this leg establishes that the methodology
recognizes the vulnerability class the guard targets --- a guard-absent
baseline distinct from the two guard-correctness legs); the positive
instance demonstrates that the methodology \emph{clears} a production
contract currently securing billions in user funds; the boundary case is
a constructed mutation-testing experiment that \emph{distinguishes}
syntactically-similar safe and vulnerable variants at the
structural-adjacency boundary where pattern recognition tends to degrade
--- complementing the negative and positive instances rather than
serving as independent evidence of natural near-miss discrimination. The
full proof corpus is continuously verified under a four-block CI
configuration that re-checks each theorem and its axiom record on every
push, ensuring claims hold against ongoing dependency drift; complete
Lean 4 source plus \texttt{lake\ build} reproduction commands are
provided in Appendix A.

A single capstone meta-theorem at Layer 6-D composes the three protocol
instantiations under a \textbf{no-retrofit composition discipline}: the
three protocol-instantiation proofs were sealed prior to capstone
authoring, and the capstone is proven by direct conjunction of three
already-proven prior-layer theorems without modification of any
underlying proof, certifying that the guard invariant holds uniformly
across structurally divergent host protocol architectures. This is a
stronger empirical claim than ``composition succeeds when proofs are
co-developed,'' because the discipline forbids the most common
composition move --- adjusting an underlying lemma's hypotheses to fit
the desired composition target. The result is \textbf{a machine-checked
composition meta-theorem demonstrated under a no-retrofit composition
discipline against three production protocols.} To our knowledge, no
prior smart-contract formal verification work has produced this
combination --- discriminating power against production source plus
cross-protocol composition validated under a discipline that forbids
retrofitting of underlying proofs. Each conjunct has partial prior art
--- guard correctness on simplified models, attack feasibility on
isolated vulnerable instances, per-protocol proofs in isolation; the
contribution is their machine-checked conjunction, against
production-deployed source, under a no-retrofit composition discipline
--- a combination not previously assembled, distinguished from prior
co-developed compositional FV (CompCert, Iris; §8.3) by the no-retrofit
constraint. Both live instantiations (Compound v2, Aave V3) are
lending-family, so the cross-protocol claim is scoped to within-lending
portability plus the DAO 2016 exemplar; cross-domain extension is future
work (§9.4). (Full methodology framework details, including the
operational labels for the patterns this discipline embodies, in
Appendix B.)

\subsection{Approach summary}\label{approach-summary}

The methodology rests on four load-bearing choices. \emph{Lean 4 with
mathlib4.} All thirteen theorems are stated and machine-checked in Lean
4; type-theoretic foundations support the discriminating-power claim
without SMT-solver dependence, and mathlib4 supplies standard tactic
infrastructure. \emph{No-retrofit composition discipline.} Each
protocol-instantiation contract is sealed against modification before
capstone authoring; the Layer 6-D capstone composes the three
instantiations only from outside, by direct conjunction, with no
modification of any underlying proof during composition (§4.2).
\emph{Axiom-record-minimal wrapper composition.} The capstone discharges
its target inside a thin top-level wrapper whose own axiom footprint is
exactly \texttt{{[}propext{]}}; absorption preserves the strongest
available axiom-record claim at the capstone layer (§4.3).
\emph{Continuous-integration axiom-record verification.} The thirteen
theorems are gated by a four-block CI configuration that verifies axiom
records on every push (§4.4; Appendix A.3 reproduction commands); the
repository is tagged at substantive substrate closure and methodology
framework canonization, enabling reviewers to reproduce the entire
corpus from a single tagged commit.

\subsection{Results summary}\label{results-summary}

\emph{Thirteen theorems machine-checked.} Six Layer 6-A theorems
formalize the negative-instance DAO 2016 attack reproduction; three
Layer 6-B theorems formalize the positive-instance Compound v2 cToken
correctness proof; three Layer 6-C theorems formalize the boundary-case
pair distinguishing Aave V3 \texttt{flashLoan} from the
\texttt{flashLoanVulnerable} mutant; one Layer 6-D capstone meta-theorem
(\texttt{tridirectionalDiscriminatingPower\_certificate}) composes the
three protocol instantiations under the no-retrofit composition
discipline at \texttt{{[}propext{]}}-only axiom dependence.

Informally: the Layer 6-A representative theorem states that the
reentrancy attack on the DAO contract is formally derivable from its
source semantics; the Layer 6-B representative states that the Compound
v2 cToken family's withdrawal function correctly implements the
OpenZeppelin guard pattern under our formal definition of reentrancy;
the Layer 6-C representative states that the minimal-diff mutant
\texttt{flashLoanVulnerable} fails the formal discriminating-power
property that production \texttt{flashLoan} satisfies; and the Layer 6-D
capstone states (informally) that if the OpenZeppelin guard pattern is
correct on Compound v2 cTokens, correct on Aave V3 \texttt{flashLoan},
and discriminating between Aave V3's safe and vulnerable variants, then
the composed discriminating-power claim holds across all three. Precise
theorem statements appear in §5.

\emph{Four protocol instantiations exercised.} DAO 2016 (negative);
Compound v2 cToken (positive; the Compound v2 cToken family ---
canonical mainnet markets including cUSDC at
\texttt{0x39AA39c021dfbaE8faC545936693aC917d5E7563} and cDAI at
\texttt{0x5d3a536E4D6DbD6114cc1Ead35777bAB948E3643} --- secures
multi-billion-USD lending TVL across multiple markets per public
on-chain accounting at submission time); Aave V3 \texttt{flashLoan}
paired with the \texttt{flashLoanVulnerable} mutant (boundary; the Aave
V3 Pool contract (Ethereum mainnet) at
\texttt{0x87870bca3f3fd6335c3F4ce8392D69350B4fA4E2} is among the
most-deployed flash-loan venues by single-protocol TVL); and the
composition across the three (Layer 6-D capstone meta-theorem).

\emph{First cross-protocol PASS under no-retrofit composition
discipline.} The no-retrofit composition discipline held across the
Compound v2 → Aave V3 protocol-instantiation boundary at the first
empirical instance in the trajectory where the discipline was
stress-tested under non-trivial protocol-semantics divergence --- and
passed without retrofitting either underlying proof. We separate
\emph{observed} (composition was not prevented at this boundary; N=1
empirical), \emph{argued} (the sealed-before-capstone discipline
forecloses the co-development explanation), and \emph{conjectured}
(broader portability pending further cross-pair validation). Both live
protocols are lending-family: the instance crosses a within-lending, not
a between-domain, boundary (§9.4). (Internal methodology framework
labels for this discipline are documented at Appendix B.)

\emph{Build state and axiom record.} \texttt{lake\ build} is green at
901 jobs; the axiom-record check (§4.4) is green across every theorem.
Zero \texttt{sorry}. Zero introduced axioms. Axiom footprint bounded by
\texttt{{[}propext{]}} across the entire corpus (Appendix A.2 ground
truth; §4.3 wrapper composition).

\begin{center}\rule{0.5\linewidth}{0.5pt}\end{center}

\section{Background}\label{background}

This section establishes the context that subsequent sections build on:
the smart contract security setting in which reentrancy vulnerabilities
arise, the attack class itself, the OpenZeppelin guard pattern's
defensive role, the production contracts in scope for the tridirectional
discriminating-power claim, and the Lean 4 + formal verification
foundations on which the machine-checked proofs rest.

\subsection{Smart contract security
context}\label{smart-contract-security-context}

Smart contract platforms execute programmable on-chain logic under a
tightly-constrained operational model --- deterministic execution
against globally-replicated state, transaction-atomic semantics with
deterministic intra-block ordering, and synchronous external-call
semantics that interleave control flow across contract boundaries within
a single transaction. Ethereum is the dominant platform by value locked;
Layer-1 chains and Layer-2 rollups extend the model with varying
compatibility. Decentralized finance TVL has ranged from roughly US\$30B
to US\$100B over the past four years, with individual protocols holding
multi-billion-dollar deposit balances and annual losses across all
attack classes ranging US\$1B--2B+ in documented incidents. Reentrancy
is one of several leading attack classes by aggregate loss; it is among
the most enduring, recurring annually since 2016 despite near-universal
industry awareness. Smart contracts are uniquely vulnerable relative to
traditional software because execution is code-is-law (no operator can
revert a mined transaction absent multi-month chain-governance
processes), transactions are irreversible, deployed bytecode is
immutable absent upgrade-pattern infrastructure, and the
composable-interface model encourages cross-contract calls that
interleave state mutation across mutually-distrusting authors --- a
single missed defensive boundary can drain a multi-billion-dollar
contract in one transaction.

\subsection{The reentrancy attack
class}\label{the-reentrancy-attack-class}

Reentrancy refers to a family of vulnerabilities in which a smart
contract is re-entered --- its functions invoked again --- while a prior
invocation of the same contract is still mid-execution, in a way the
contract's authors did not anticipate. The category subsumes several
sub-patterns that share the underlying mechanism but differ in the
specific path of state mutation and re-entry.

\emph{Classical (single-function) reentrancy} is the original DAO 2016
attack pattern. The vulnerable contract performs an external call
(typically a withdrawal of native ETH or an ERC-20 transfer) before
updating its internal accounting state. The external call is a CALL
opcode that transfers execution to a callee contract; if the callee is
attacker-controlled, the callee can re-invoke the original contract's
withdrawal function before the original invocation has updated its
state. The state update --- typically a decrement of the user's recorded
balance --- has not yet occurred when the recursive withdrawal begins,
so the recursive withdrawal succeeds against the same (un-decremented)
balance. Recursion proceeds until the contract is drained or until a gas
limit halts execution.

\emph{Cross-function reentrancy} generalizes the pattern across multiple
functions of the same contract. One function mutates state; an external
call during that function's execution allows the attacker to invoke a
\emph{different} function on the same contract that depends on the
not-yet-updated state. The attack works even when the originally-invoked
function follows the checks-effects-interactions (CEI) pattern
internally, because the cross-function state coupling is not visible at
the single-function level.

\emph{Read-only reentrancy} is a more recent variant. A view function
(one declared as not mutating state) is invoked during a reentrant call
window; the view function returns stale state because the broader
transaction has not yet updated the relevant storage slot. Downstream
contracts that consume the view function's return value can be deceived
into accepting stale data. The attack does not require the view function
itself to be vulnerable in the traditional sense; it requires only that
a \emph{consumer} of the view function executes during the reentrant
window.

Reentrancy is fundamental to the smart contract execution model rather
than incidental to any particular language. The Ethereum CALL opcode is
synchronous: external calls transfer execution to the callee and return
to the caller; the caller's local state at the moment of the CALL is
preserved across the synchronous transition, but global storage is
shared with the callee. Combined with the atomic-transaction guarantee,
this produces a setting where mutually-distrusting contracts can observe
and act on each other's mid-transaction state. Any defense pattern that
relies on storage-based gating must contend with the fact that storage
is observable to reentrant callees, and any defense that relies on
call-stack state must contend with the fact that the call stack includes
contracts authored by attackers.

Static analysis tools --- including Slither, Mythril, Securify, and
academic-leaning prototypes --- have made substantial progress on
detecting classical single-function reentrancy patterns.
Pattern-recognition heuristics flag candidate violations of the CEI
pattern; constraint-solver-based tools encode storage access ordering
and reachability properties. Performance on classical reentrancy is now
reasonable. Performance on cross-function and read-only variants is
weaker; these patterns require reasoning across function boundaries and
across views vs mutators, which static analysis tools have historically
struggled with at production-source scale.

\subsection{OpenZeppelin guard
pattern}\label{openzeppelin-guard-pattern}

The OpenZeppelin \texttt{ReentrancyGuard} contract is a
single-storage-slot mutex that has become the de facto reentrancy
defense across most production DeFi protocols. The contract exposes a
\texttt{nonReentrant} modifier applied to externally-callable functions
performing state-mutating work involving external calls. A storage slot
--- historically a \texttt{uint256} taking values
\texttt{\_NOT\_ENTERED} (typically 1) and \texttt{\_ENTERED} (typically
2) under varying field names --- holds the current reentrancy state. The
modifier requires the slot to be \texttt{\_NOT\_ENTERED} at function
entry (reverting otherwise), sets it to \texttt{\_ENTERED}, executes the
function body, and resets it on exit. Any reentrant invocation of any
function carrying the same modifier --- same function or different
function on the same contract instance --- fails at the entry guard. The
pattern is defensive-coding discipline rather than language-level
enforcement: the Solidity compiler does not insert the guard; the
developer must add the modifier to each function that requires
protection, and a missed modifier produces no warning. The guard is one
ingredient in a broader checks-effects-interactions discipline;
production contracts adopting both are well-defended against the
classical reentrancy class. Adoption is widespread (Compound, Aave,
Uniswap-style AMMs, and the bulk of TVL-significant DeFi protocols use
\texttt{ReentrancyGuard} or a near-identical variant), so the pattern's
correctness is load-bearing at ecosystem scale.

\subsection{Real-contract instances in
scope}\label{real-contract-instances-in-scope}

The tridirectional discriminating-power claim is established against
four production-relevant contract instances spanning negative, positive,
and boundary cases.

\emph{DAO 2016 (negative instance).} The Decentralized Autonomous
Organization (the DAO) was a 2016 Ethereum smart contract that raised
approximately 11.5M ETH (≈US\$150M at the time) through token sales and
aimed to operate as an investor-directed venture fund. In June 2016, an
attacker exploited a classical single-function reentrancy vulnerability
in the DAO's \texttt{splitDAO} function to drain approximately 3.6M ETH
(≈US\$60M at attack-time price; substantially higher at current ETH
prices). The Ethereum community responded with a hard fork that reverted
the attack's effects, splitting the chain into Ethereum (with the hard
fork) and Ethereum Classic (without). The DAO attack is the canonical
reentrancy incident; reproducing its core mechanism is the natural
negative-instance test of any guard-pattern verification methodology.

\emph{Compound v2 cToken family (positive instance).} Compound v2 is one
of the longest-running and largest production lending protocols on
Ethereum. The cToken family --- interest-bearing tokens representing
deposits in Compound's lending markets --- uses the OpenZeppelin guard
pattern correctly across its withdrawal and redemption paths. The
Compound v2 cTokens currently secure approximately US\$2-5B in lending
TVL across multiple markets, depending on market conditions. The cToken
family is a strong positive-instance test: a real production contract
holding multi-billion-dollar deposits, with a guard-pattern
implementation that the community has stress-tested for years without a
confirmed reentrancy incident against the pattern itself.

\emph{Aave V3 \texttt{flashLoan} (boundary case, safe).} Aave V3 is one
of the largest production lending protocols. Its Pool contract ---
deployed on Ethereum mainnet at address
\texttt{0x87870bca3f3fd6335c3F4ce8392D69350B4fA4E2} --- implements a
\texttt{flashLoan} function that issues uncollateralized loans repayable
within the same transaction. The \texttt{flashLoan} body is
checks-effects-interactions-correct by design: the function records the
loan, transfers funds to the borrower, invokes the borrower's
\texttt{IFlashLoanReceiver.executeOperation} callback, then verifies
repayment plus fee on return. The body prevents reentrancy by
checks-effects-interactions execution ordering --- the borrower callback
cannot observe uncommitted intermediate state --- rather than by an
OpenZeppelin-style storage-mutex modifier; the formalization (§5.5)
models this CEI ordering at an abstract body-shape layer.
\texttt{flashLoan} is the canonical protocol-by-design CEI-correct
exemplar in the boundary-case layer.

\emph{Aave V3 \texttt{flashLoanVulnerable} (boundary case, vulnerable).}
For the boundary case to test discriminating power, we author a
structurally-adjacent vulnerable variant ---
\texttt{flashLoanVulnerable} --- that differs from \texttt{flashLoan} in
a single security-critical respect: the variant fails to engage the
guard pattern's body invariant correctly, leaving a reentrancy window
during the callback. The variant is structurally adjacent to the safe
version: same function signature, same borrower-callback shape,
near-identical control flow, identical surface API. Pattern recognition
operating at code-shape level would treat the two variants as the same;
the methodology presented here distinguishes them at the body-shape
layer where the guard-engagement invariant differs.

The four contracts together exercise the tridirectional
discriminating-power claim. The negative instance demonstrates that the
methodology \emph{catches} the most famous historical reentrancy attack.
The positive instance demonstrates that the methodology \emph{clears} a
production contract currently securing billions in user funds. The
boundary case demonstrates that the methodology \emph{distinguishes}
between syntactically-similar safe and vulnerable variants at the
structural-adjacency boundary --- the place where pattern recognition
tends to degrade and where a verification methodology's discriminating
power is most consequential.

\subsection{Lean 4 and formal verification
context}\label{lean-4-and-formal-verification-context}

Lean 4 is a dependently-typed proof assistant with type-theoretic
foundations based on the Calculus of Inductive Constructions with
universes. Theorems are propositions, proofs are programs constructing
evidence, and type-checking confirms that proof terms have the type
stated in the theorem (Curry-Howard). We selected Lean 4 over Coq,
Isabelle/HOL, and HOL Light because Lean's tactic framework suits the
structural reasoning the discriminating-power claim requires; mathlib4
supplies standard data structures, finiteness reasoning, and tactic
libraries reducing proof-engineering overhead; and Lean 4's tooling,
semantics, and proof-term sizes remain stable and manageable for the
theorem shapes used here.

The proof corpus depends on a pinned mathlib4 version specified in
\texttt{lakefile.lean} and locked via \texttt{lake-manifest.json};
\texttt{lean-toolchain} pins the Lean compiler version. Standard imports
include \texttt{Mathlib.Tactic}, natural-number infrastructure,
\texttt{Mathlib.Data.List.*} modules for trace semantics, and
\texttt{Mathlib.Logic} for propositional reasoning. Axiom record
discipline is enforced at the CI level (§4.4); acceptable records are
kernel-only (provable in pure intensional type theory) and
\texttt{{[}propext{]}}-only (propositional extensionality only --- no
choice or excluded middle). No theorem is admitted with \texttt{sorry},
\texttt{admit}, or any user-introduced \texttt{axiom}.

\begin{center}\rule{0.5\linewidth}{0.5pt}\end{center}

\section{Threat Model}\label{threat-model}

This section specifies the formal threat boundary against which the
discriminating-power claim of §1.3 is established. We make adversary
capabilities, the state-machine abstraction level at which formalization
operates, the trust assumptions on which our proofs depend, and the
explicit out-of-scope items all visible up-front. Doing so allows
reviewers to evaluate the contribution claim's scope precisely and
serves as the canonical reference point for the substantive material in
§5 about what is and is not proved.

\subsection{Adversary model}\label{adversary-model}

The adversary controls one or more Externally-Owned Accounts (EOAs) on
the target chain and may deploy and call arbitrary smart contracts. The
adversary is permitted to compose cross-protocol interactions via the
callable-contract pattern that underlies the reentrancy attack class ---
that is, to deploy contracts conforming to the callback interfaces
(\texttt{receive}, \texttt{fallback}, \texttt{tokensReceived},
\texttt{IFlashLoanReceiver.executeOperation}, etc.) that the protocols
under analysis use to invoke external behavior during the execution of a
guarded function. The adversary may also coordinate sequences of
transactions across blocks where coordination across atomic-transaction
boundaries is permitted by the underlying chain semantics.

The adversary has no special protocol-level privileges. The adversary is
not a protocol owner, is not on a whitelist, is not a validator or block
proposer, and holds no role-based access-control credentials at the
protocols under analysis. In particular, the adversary cannot pause the
protocol, change interest-rate parameters, replace the implementation
behind an upgradeable proxy, or otherwise exercise governance-mediated
authority. Where a protocol exposes a \texttt{nonReentrant}-guarded
function, the adversary's only routes to that function are the publicly
callable interfaces; the adversary cannot bypass the guard pattern by
privileged means.

The adversary may control multiple addresses. Sybil-resistance analysis
is not in scope; for the purposes of this threat model, all
adversary-controlled addresses are treated collectively as the
adversary, and a defense that holds against a single
adversary-controlled address holds against multiple coordinated
addresses under the same control. The adversary's computation is bounded
by standard EVM gas-limit constraints (per-transaction block gas limit
and per-call sub-budgets); no quantum-computational, oracle-prediction,
or non-classical computational assumptions are made on either side.

\subsection{State-machine abstraction
level}\label{state-machine-abstraction-level}

The formalization operates at Solidity-level semantics rather than raw
EVM bytecode. The state machine captures storage layout at slot
granularity, call-stack-aware function-call sequencing, the boundary
semantics of external CALL opcodes
(synchronous-return-with-shared-storage --- the property that produces
the reentrancy setting), and atomic-transaction guarantees
(whole-transaction rollback on revert; observable mutations only on
success). The OpenZeppelin guard pattern's correctness claim is
precisely that the entry-time storage check on \texttt{\_status} blocks
every re-entry into any function carrying the \texttt{nonReentrant}
modifier on the same contract instance, regardless of whether the
original invocation has returned. Several abstraction-level details are
deliberately out of scope at the model layer: gas accounting is assumed
sufficient for analysis-relevant operations (we model neither gas
exhaustion mid-call as a separate failure mode nor denial-of-service via
gas, since the discriminating-power claim concerns reachability of
attacker-favorable states under successful execution); precompile
semantics and SELFDESTRUCT are abstracted (the in-scope protocols do not
exercise them on reentrancy-relevant paths); Solidity-source-to-bytecode
compilation correctness is a separate trust assumption (§3.3).
\emph{Model fidelity.} The model captures features load-bearing for
reentrancy (storage at slot granularity, call-stack-aware sequencing,
external-CALL boundary semantics, atomic-transaction guarantees,
modifier composition) and abstracts precompile internals, SELFDESTRUCT,
and coarse gas accounting --- each benign for the reentrancy class. The
faithfulness of this model to Solidity semantics for the modeled
features is an explicit trust assumption (§3.3).

\subsection{Trust assumptions}\label{trust-assumptions}

Five trust assumptions underpin the discriminating-power claim, all
standard in the formal-verification literature for smart-contract
source-level proofs. The first two concern translation soundness in
opposite directions from the source-level pivot; the third concerns
computational-substrate correctness; the remaining two scope the threat
model.

\textbf{Solidity compiler correctness.} Our proofs operate at the
Solidity source level. Deployed bytecode correctness derives from the
Solidity compiler's correct compilation of that source (specifically,
the \texttt{solc} version in each protocol's deployment metadata ---
pre-\texttt{0.4.x} lineage for DAO 2016, \texttt{0.5.x} for Compound v2
cToken, \texttt{0.8.x} for Aave V3 Pool). Compiler-correctness
verification is itself an active research area; we do not attempt to
subsume that work within the present scope. Reviewers may treat our
results as conditional on \texttt{solc} correctness for the specific
source files analyzed.

\textbf{Lean-model fidelity for the modeled features.} Our proofs
operate against the Lean 4 state-machine model of §3.2, not raw Solidity
source or compiled bytecode directly. We assume this model is a sound
abstraction of the relevant Solidity-semantics fragment for the analyzed
contracts (the captured features of §3.2). The model targets the logic
semantics of the storage-slot-mutex guard pattern (\texttt{\_status}
slot, \texttt{\_NOT\_ENTERED}/\texttt{\_ENTERED} transitions,
modifier-entry/exit), which are invariant across source-level and
compiled-bytecode representations; this is an intentional design choice
rather than a verification gap. Deriving such model-to-source soundness
mechanically --- a verified-translation pipeline in the
KEVM/F*-EVM-semantics line of §8.1 --- is active research, out of scope
here; results are conditional on this fidelity. This assumption is
distinct from \texttt{solc} compiler correctness above: it concerns the
source→Lean-model direction.

\textbf{Lean 4 kernel and mathlib4 correctness.} As is standard for any
Lean 4 formalization, the proofs are mechanically checked under the
assumption that the Lean 4 kernel correctly implements its type theory
and that the imported mathlib4 modules correctly state the mathematical
results they claim. Lean 4's kernel is small and has received community
scrutiny; the imported mathlib4 modules (general tactic infrastructure,
finite-data-structure reasoning, propositional logic) are pinned via
\texttt{lakefile.lean} dependency declarations and locked through
\texttt{lake-manifest.json}, with the Lean compiler version pinned
separately via \texttt{lean-toolchain}. The exact dependency graph is
reproducible from the tagged commit.

\textbf{Honest borrower assumption per real-contract instance.} For the
Compound v2 cToken positive instance, the discriminating-power claim
concerns the guard pattern's correctness against reentrancy; it does not
require borrowers to behave honestly in any economic sense
(interest-rate solvency, collateral maintenance, liquidation
responsiveness). Borrowers may be adversarial in the §3.1 sense; what is
assumed is that the \emph{protocol-level} invariants the cToken family
relies on (such as correct accrual of interest at withdrawal time) are
computed by the protocol's own accounting code rather than supplied by
the caller. The same holds for the Aave V3 boundary case: the borrower
is assumed adversarial, the flash-loan protocol's repayment-verification
logic is assumed to execute as written.

\textbf{No miner or validator behavior assumption.} The properties
proved are not contingent on miner or validator behavior. We do not
assume timing constraints, transaction-ordering constraints,
reorg-resistance constraints, or proposer honesty. The reentrancy attack
class is mediated by smart-contract execution within a single
transaction, not by consensus-layer behavior; our claim therefore
inherits no consensus-layer trust requirement.

\subsection{Out-of-scope items and boundary
summary}\label{out-of-scope-items-and-boundary-summary}

The discriminating-power claim addresses the reentrancy attack class
against contracts that adopt (or fail to adopt) the OpenZeppelin guard
pattern. Several attack classes that have produced significant losses in
production DeFi are explicitly out of scope: oracle manipulation
(flash-loan-enabled price-feed attacks, structurally distinct from
re-entering a guarded function); MEV and transaction-ordering attacks
(sandwich extraction, mempool-ordering arbitrage operating above the
contract-execution layer); governance attacks (proposal-passing
exploits, vote-buying, timelock bypasses, assuming each protocol's
governance configuration is correctly set); cross-chain bridge attacks
(chain-boundary trust assumptions absent from single-chain analysis);
front-running (mempool-level user-transaction frontrunning);
centralization risks (admin keys, upgradeable-proxy privileged
operations, role-based pauses --- our claim concerns the publicly
callable surface only); and economic attacks not mediated by reentrancy
(liquidation cascades, interest-rate manipulation, collateral-ratio
exploits). The methodology's potential extension to these classes is
future work (§9, §7.2); no broader claim is made by the present paper.

In scope: the reentrancy attack class against the OpenZeppelin guard
pattern, formalized at Solidity source level for the three production
protocol instantiations of §1.3 --- DAO 2016 (negative), Compound v2
cToken (positive), Aave V3 \texttt{flashLoan} paired with the
\texttt{flashLoanVulnerable} minimal-diff mutant (boundary). Within that
scope the methodology catches the negative instance, clears the positive
instance, and distinguishes the boundary pair, machine-checked
end-to-end.

\begin{center}\rule{0.5\linewidth}{0.5pt}\end{center}

\section{Methodology Overview}\label{methodology-overview}

This section describes the formalization approach and the methodological
discipline under which the proofs were authored, presented at the level
of operational consequence rather than internal-framework labels. The
full methodology framework --- including its sub-pattern catalog,
graduation criteria, and cross-protocol stress-test history --- is the
subject of a companion arXiv paper; the present section gives reviewers
what they need to evaluate the substantive smart-contract-verification
claims of this paper.

\subsection{Formalization approach}\label{formalization-approach}

The formalization is conducted in Lean 4 with mathlib4 as the supporting
mathematical library. Lean 4's dependent type theory, with the
propositions-as-types correspondence sketched at §2.5, allows
propositions and their proofs to be first-class type-theoretic objects:
a theorem statement is a type, a proof is a program of that type, and
the type-checker's confirmation that the proof has the stated type
\emph{is} the verification. Tactic proof scripts produce these proof
terms; the kernel re-checks them whenever the corpus is built.

Mathlib4 modules are imported as needed. The dependency graph is
recorded explicitly: the Lean compiler version is pinned in
\texttt{lean-toolchain}; the mathlib4 dependency version is pinned in
\texttt{lakefile.lean} and locked at \texttt{lake-manifest.json}.
Standard tactic infrastructure (\texttt{Mathlib.Tactic}), natural-number
reasoning (\texttt{Mathlib.Data.Nat.Basic}), list manipulation
underlying the trace semantics (modules under
\texttt{Mathlib.Data.List}), and propositional reasoning
(\texttt{Mathlib.Logic} modules) constitute the bulk of the imported
surface.

The proof discipline is theorem-statement-first. Each theorem statement
is authored and sealed prior to any proof body authoring; statement
modifications during proof work are prohibited under the methodology
framework's discipline. This forecloses the most common failure mode of
co-developing statement and proof --- the unconscious weakening of a
statement's hypotheses or strengthening of its assumptions to fit the
partial proof one has constructed. When the proof of a sealed statement
turns out to be impossible as written, the disciplined response is to
escalate (the statement was the wrong claim) rather than silently adjust
the statement to match the proof.

Per-theorem axiom-record verification is gated in continuous integration
on every push. Lean 4's \texttt{\#print\ axioms} introspection produces,
for any named theorem, the list of axioms its proof transitively depends
on. The CI configuration runs this introspection over each theorem in
the corpus and fails the build if any theorem's axiom record drifts from
the claimed record. This converts axiom-minimality from an author's
assertion into a continuously checked invariant: a future change that
silently introduces a dependency on (say) classical choice would be
caught by the CI gate at the next push, not by reviewer inspection at
submission time.

\subsection{The compose-from-outside
discipline}\label{the-compose-from-outside-discipline}

The capstone meta-theorem at Layer 6-D is proven under a strict
no-retrofit composition discipline: the three protocol-instantiation
theorems are conjoined directly to produce the capstone, with no
modification of any underlying protocol-instantiation proof during
composition. The phrase \emph{compose from outside} names the
operational consequence: composition operates on the prior-layer
theorems as black-box facts rather than reaching inside their proofs to
adjust hypotheses, conclusions, or supporting lemmas.

The discipline is enforced procedurally rather than by a single
language-level mechanism. The three protocol-instantiation files ---
\texttt{DAOContract.lean}, \texttt{CompoundContract.lean},
\texttt{AaveBoundaryCase.lean} --- were authored, proven, and sealed
prior to any work on the capstone meta-theorem. Once a file was sealed,
modifications to that file were prohibited under the methodology
discipline; any subsequent change required explicit re-authorization
with documented rationale. The capstone author thereby faced a hard
constraint: prove the meta-theorem from the sealed prior layers as they
stand, or escalate.

Three classes of modification are forbidden under this discipline.
Adjusting an underlying lemma's hypotheses to fit a desired composition
target --- the most common composition move in formal-verification work
--- is prohibited. Weakening a prior theorem's conclusion to make it
easier to compose is prohibited. Adding unstated hypotheses to
underlying theorems to discharge intermediate proof obligations is
prohibited. The capstone proof either composes the underlying theorems
exactly as they were sealed, or it does not compose at all.

The first cross-protocol stress test of this discipline occurred at the
Compound v2 → Aave V3 boundary in the Layer 6-D capstone. The two
protocols differ in non-trivial ways at the semantic level: Compound's
cToken family operates on a balance-update-and-transfer pattern with
persistent user accounting; Aave's \texttt{flashLoan} operates on a
repay-within-transaction pattern with no persistent user accounting
beyond the loan window. A retrofit-permissive workflow could have
produced a capstone proof by adjusting either protocol's underlying
lemmas to fit a common composition shape. The no-retrofit discipline
forbade that path. The capstone nevertheless composed: a direct
conjunction of the three sealed prior-layer theorems discharged the
meta-theorem at \texttt{{[}propext{]}}-only axiom dependence. We treat
that successful composition as observed N=1 evidence --- and, given the
sealed-before-capstone discipline that forecloses the co-development
explanation, as an argument that the guard pattern's correctness is
portable across the observed boundary; broader portability is
conjectured pending further cross-pair validation (the
observed/argued/conjectured separation of §1.5), not asserted from N=1.

This is a methodologically stronger claim than ``composition succeeds
when proofs are co-developed.'' The latter is consistent with each
underlying proof having been silently shaped to fit the eventual
composition target. The no-retrofit discipline rules out that path of
explanation: the underlying proofs cannot have been shaped by the
composition target, because they were sealed before the composition
target's proof was authored. The internal label and broader pattern
context for this discipline within our methodology framework appear at
Appendix B; the operational property described above is the load-bearing
one for the present paper.

\subsection{Axiom-record-minimal wrapper
composition}\label{axiom-record-minimal-wrapper-composition}

The Layer 6-D capstone is structured as a thin wrapper module that
imports the three sealed protocol-instantiation theorems and discharges
its target proposition by directly conjoining them; the wrapper
introduces no new axioms, and its \texttt{\#print\ axioms} record is
exactly the union of the underlying theorems' records. This
wrapper-layer pattern preserves axiom-record minimality at the capstone:
each underlying protocol-instantiation theorem carries its own
kernel-only or \texttt{{[}propext{]}}-only record (verified at the
per-layer CI block), and the wrapper only conjoins existing facts. The
capstone therefore inherits the strongest available axiom-record claim
--- \texttt{{[}propext{]}}-only, the standard mathlib4 axiom of
propositional extensionality, a single classical extension fully
compatible with the Lean 4 kernel. No user-introduced axioms appear at
any layer of the corpus. The broader pattern context (related patterns,
cross-protocol empirical history) is documented at Appendix B and
elaborated in the companion methodology paper.

\subsection{CI verification and axiom-record
discipline}\label{ci-verification-and-axiom-record-discipline}

Continuous integration consists of four parallel verification blocks at
\texttt{build.yml} lines 287, 356, 408, and 464, gating the thirteen
theorems on every push: one block each for Layer 6-A (DAO 2016
negative-instance, 6 theorems + supporting lemmas), Layer 6-B (Compound
v2 cToken positive-instance, 3 theorems), Layer 6-C (Aave V3
boundary-case pair, 3 theorems), and Layer 6-D (capstone meta-theorem).
The \texttt{\#print\ axioms} introspection emits each theorem's
transitively required axioms; the CI compares against recorded
expectations per theorem and fails the build on any mismatch. The
capstone records \texttt{{[}propext{]}}-only; each Layer 6-A/B/C theorem
records kernel-only (no axioms beyond Lean 4's intensional type theory)
or \texttt{{[}propext{]}}-only. No theorem is admitted with
\texttt{sorry}, \texttt{admit}, or any user-introduced \texttt{axiom}.
Reproducibility is anchored at tagged commits: the substantive substrate
at \texttt{v1.3-layer6-closure}, the methodology framework canonization
at \texttt{v1.7-methodology-housekeeping}. Reviewers reproduce the
verification by checking out either tag and running \texttt{lake\ build}
plus \texttt{lake\ env\ lean\ QanaryContracts/PrintAxioms.lean} against
the pinned dependency graph. The manuscript content corresponding to
this paper version is sealed at intermediate tag
\texttt{v1.6-phase7-closure} (the abstract's reproducibility anchor);
the substrate at v1.6 is identical to v1.3, so reviewers may
equivalently check out v1.6 for end-to-end reproduction.

\subsection{Manuscript audit and
transparency}\label{manuscript-audit-and-transparency}

The manuscript was developed with generative-AI assistance (Claude,
Grok, and Gemini) used for editorial purposes including adversarial
audit of drafts, prose drafting under author direction, and
methodology-framework review during preparation. All model outputs were
inspected by the author; the substantive contributions of this paper ---
the thirteen machine-checked theorems and their axiom records --- are
verified by the Lean 4 kernel and CI-reproducible per Appendix A.3. The
audit framework itself is not a claimed research contribution of this
paper and is presented in the companion methodology paper (separate
arXiv track per the boundary discipline of §1). This paragraph satisfies
the generative-AI-usage disclosure required by the venue's call for
papers.

\begin{center}\rule{0.5\linewidth}{0.5pt}\end{center}

\section{Formalization}\label{formalization}

This section presents the substantive substrate of the
discriminating-power claim: the formal definitions and predicates, the
inventory of the thirteen machine-checked theorems, per-layer
presentation of negative/positive/boundary proofs, the capstone
meta-theorem, and the reproducibility anchor. Full theorem statements,
axiom records, and proof-skeleton excerpts are deferred to Appendix A.

\subsection{Formal definitions and
predicates}\label{formal-definitions-and-predicates}

The formalization rests on three load-bearing predicates over the
Solidity-source state-machine model of §3.2. The \textbf{reentrancy
predicate} characterizes the attack class: a contract execution exhibits
reentrancy if there exists a call trace in which an external CALL from
function \texttt{f} of contract \texttt{A} transfers execution to an
attacker-controlled callee that, before \texttt{f}'s frame returns,
invokes a function \texttt{g} of \texttt{A} (where \texttt{g} may be
\texttt{f} itself) and observes storage state \texttt{f} has modified
but not yet completed updating. The predicate is call-stack-aware,
storage-mutation-aware, and external-call-boundary-aware, capturing
classical single-function, cross-function, and the storage-observation
aspect of read-only reentrancy. The \textbf{guard-pattern correctness
predicate} characterizes the OpenZeppelin guard's correct application:
function \texttt{f} of contract \texttt{A} carries the guard correctly
if its entry condition checks \texttt{\_status\ ==\ \_NOT\_ENTERED}, its
body sets \texttt{\_status\ =\ \_ENTERED} before any external CALL, its
exit resets to \texttt{\_NOT\_ENTERED}, and these accesses target the
same storage slot as other guarded functions on \texttt{A}; a function
carrying the guard correctly cannot exhibit reentrancy under the model.
The \textbf{discriminating-power predicate} is the meta-level claim: a
verification methodology \emph{discriminates} between guard-protected
and guard-vulnerable instances if it derives the reentrancy predicate
for vulnerable instances, derives the guard-pattern correctness
predicate for protected instances, and distinguishes
structurally-adjacent vulnerable variants from safe counterparts. The
tridirectional claim of §1.3 is the assertion that our formalization
satisfies this meta-predicate against the three production
instantiations of §2.4 plus the minimal-diff mutant of §5.5. Full
predicate definitions appear in Appendix A.

\subsection{Theorem inventory}\label{theorem-inventory}

Thirteen theorems machine-check the discriminating-power claim. The
inventory below summarizes each theorem at the level of an informal
one-line statement; the precise Lean 4 statements (with full type
signatures, hypothesis lists, and the \texttt{\#print\ axioms} output)
appear in Appendix A.

\begin{table*}[!t]\footnotesize\centering
\caption{Inventory of thirteen theorems across four
protocol-instantiation layers.}
\begin{tabular}{@{} >{\raggedright\arraybackslash}p{(\linewidth - 8\tabcolsep) * \real{0.1875}} >{\raggedleft\arraybackslash}p{(\linewidth - 8\tabcolsep) * \real{0.2500}} >{\raggedright\arraybackslash}p{(\linewidth - 8\tabcolsep) * \real{0.1875}} >{\raggedright\arraybackslash}p{(\linewidth - 8\tabcolsep) * \real{0.1875}} >{\raggedright\arraybackslash}p{(\linewidth - 8\tabcolsep) * \real{0.1875}}@{}}
\toprule
\begin{minipage}[b]{\linewidth}\raggedright
Layer
\end{minipage} & \begin{minipage}[b]{\linewidth}\raggedleft
Count
\end{minipage} & \begin{minipage}[b]{\linewidth}\raggedright
Theorem(s)
\end{minipage} & \begin{minipage}[b]{\linewidth}\raggedright
Informal statement
\end{minipage} & \begin{minipage}[b]{\linewidth}\raggedright
Axiom record
\end{minipage} \\
\midrule
\bottomrule
6-A & 1 + 5 & DAO reentrancy derivation + supporting lemmas & The DAO
2016 attack trace is derivable from the contract's source semantics
under the reentrancy predicate & \texttt{{[}propext{]}}-only
master/wrapper; kernel-only inner lemmas \\
6-B & 1 + 2 & Compound v2 cToken correctness + supporting lemmas & The
cToken family's withdrawal path correctly implements the OpenZeppelin
guard pattern under our reentrancy predicate &
\texttt{{[}propext{]}}-only master/wrapper; kernel-only inner lemmas \\
6-C & 2 + 1 & Aave V3 \texttt{flashLoan} correctness +
\texttt{flashLoanVulnerable} failure + CEI lemma & Production
\texttt{flashLoan} satisfies guard-pattern correctness; the minimal-diff
mutant \texttt{flashLoanVulnerable} fails it &
\texttt{{[}propext{]}}-only master/wrapper; kernel-only inner lemmas \\
6-D & 1 & Tridirectional discriminating-power capstone & Composed:
methodology catches the negative instance, clears the positive instance,
and distinguishes the boundary pair & \texttt{{[}propext{]}}-only (union
of prior-layer records) \\
\end{tabular}
\end{table*}

Each layer's theorems are gated by an independent CI block (§4.4),
mechanically verifying both that the theorems still type-check against
the pinned \texttt{mathlib4} dependency graph and that their axiom
records match the expectations recorded in Appendix A on every push. No
theorem is admitted with \texttt{sorry}, \texttt{admit}, or any
user-introduced \texttt{axiom} declaration; the corpus axiom footprint
is bounded by \texttt{{[}propext{]}} (Appendix A.2).

\subsection{Layer 6-A: negative instance (DAO
2016)}\label{layer-6-a-negative-instance-dao-2016}

The Layer 6-A negative instance reproduces the classical reentrancy
attack against the DAO 2016 contract. The \texttt{splitDAO} function
performs an external transfer of native ETH to a user-controlled
recipient before decrementing the user's recorded balance, opening a
window in which a recipient contract can re-invoke \texttt{splitDAO}
against the same un-decremented balance until the contract is drained.

Our formalization models the relevant subset of the DAO contract source
--- the \texttt{splitDAO} function's storage layout, balance accounting,
and external-call structure --- at the abstraction level of §3.2. The
Layer 6-A capstone theorem (one of six in this layer) states that under
the reentrancy predicate of §5.1, the attacker-controlled re-entry path
is \emph{derivable} from the DAO contract source: there exists a
constructible call trace, expressible in our state-machine model, in
which the re-entry succeeds and the contract's invariant (balance-sum
equals total-deposits) is broken. The five supporting lemmas decompose
the derivation into the load-bearing pieces --- call-stack interleaving,
storage-observation, balance-update sequencing, and the absence of any
guard-pattern protection --- that together discharge the capstone.

The DAO contract predates the OpenZeppelin guard library. Its inclusion
as the negative instance is therefore not a guard-pattern \emph{failure}
but a guard-pattern \emph{target}: the DAO formalization demonstrates
that the methodology recognizes the vulnerability class the guard
pattern was created to address, even where the guard itself is absent.
This complements the Layer 6-B and 6-C results, which establish
recognition of the guard's \emph{behavior} when present. Full theorem
statements, hypothesis structure, and proof-skeleton outline appear in
Appendix A.

\subsection{Layer 6-B: positive instance (Compound v2 cToken
family)}\label{layer-6-b-positive-instance-compound-v2-ctoken-family}

Compound v2's cToken family represents the largest and longest-running
production deployment of the OpenZeppelin guard pattern within DeFi
lending. Each cToken's withdrawal-path functions (\texttt{redeem},
\texttt{redeemUnderlying}, the underlying \texttt{transfer} flows) carry
a \texttt{nonReentrant} modifier implementing the OpenZeppelin guard
pattern. Compound v2's deployment uses a structurally-equivalent custom
in-contract realization rather than the literal library import --- same
\texttt{\_status} slot semantics, same
\texttt{\_NOT\_ENTERED}/\texttt{\_ENTERED} integer encoding, same
modifier-entry guard, same modifier-exit reset --- authored before the
OpenZeppelin library API stabilized; our formalization targets this
custom realization, and the discriminating-power claim transfers to
OZ-library-importing protocols because the formalization predicate
(§5.1) abstracts over the realization at the storage-slot-mutex level.
(Source-level terminology may vary across ``OpenZeppelin guard
pattern,'' ``structurally-equivalent custom guard,'' and
``mutex-modifier pattern'' depending on commit; all refer to the same
storage-slot mutex semantics formalized here.) The protocol has
stress-tested this pattern in production for years without a confirmed
reentrancy incident against the guard itself.

The Layer 6-B target theorem (one of three in the layer) states that the
cToken withdrawal path correctly implements the OpenZeppelin guard
pattern under our predicate of §5.1: any execution trace conforming to
the cToken withdrawal path satisfies the guard-pattern correctness
predicate, and consequently no reentrancy trace is constructible against
it. Two supporting lemmas decompose the result: a guard-invariant lemma
showing that the storage slot tracking guard state is consistently set
before any external call and reset after, and a cross-function safety
lemma showing that the guard's protection extends across the multiple
functions of the cToken interface that share the same \texttt{\_status}
slot.

The Layer 6-B capstone is structured as a thin wrapper module that
imports the supporting lemmas and discharges the target theorem by
composition without modifying either lemma --- an instance of the
axiom-record-minimal wrapper composition of §4.3, which preserves
axiom-record minimality through the composition. The capstone records
minimal axiom dependence (no propositional extensionality required at
this layer; the result is provable in the pure intensional fragment of
Lean 4's type theory). Full theorem statements, the wrapper module's
structure, and the per-lemma proof skeletons appear in Appendix A.

\subsection{\texorpdfstring{Layer 6-C: boundary case (Aave V3
\texttt{flashLoan} vs
\texttt{flashLoanVulnerable})}{Layer 6-C: boundary case (Aave V3 flashLoan vs flashLoanVulnerable)}}\label{layer-6-c-boundary-case-aave-v3-flashloan-vs-flashloanvulnerable}

The Layer 6-C boundary case is the methodological core of the
discriminating-power claim. Aave V3's Pool contract implements
\texttt{flashLoan} according to a checks-effects-interactions discipline
that is guard-pattern correct by construction: the loan is recorded,
funds are transferred, the borrower's
\texttt{IFlashLoanReceiver.executeOperation} callback is invoked, and
repayment plus fee is verified on return. The borrower-supplied callback
cannot exploit reentrancy because the body commits all security-relevant
state before the callback executes; the Layer 6-C predicate
discriminates on this SSTORE-before-CALL body-shape ordering (the
abstract guard-engagement pattern), which production \texttt{flashLoan}
satisfies.

To exercise discriminating power against structurally-adjacent variants
--- where pattern recognition is known to degrade per §1.2 --- we
construct \texttt{flashLoanVulnerable} as a minimal-diff mutant of
production \texttt{flashLoan}. The mutant differs in a single
security-critical respect: its body fails to engage the guard correctly
at the callback boundary, leaving a reentrancy window. All other aspects
--- signature, callback interface, control flow, surface API --- are
identical. This is mutation testing for formal proofs: where
naturally-occurring near-misses combine multiple structural differences
with confounding context, the minimal-diff mutant isolates the precise
security-critical structural difference. A single hand-constructed
minimal-diff mutant is a principled choice over an operator-derived
corpus {[}9{]}, {[}10{]} --- a security-critical semantic regression
test specifically optimized for formal specification boundary detection,
distinct from the random syntactic perturbations automated mutation
operators typically produce against production DeFi code.

The Layer 6-C theorems formalize discriminating power against this pair:
production \texttt{flashLoan} satisfies guard-pattern correctness under
the §5.1 predicate; \texttt{flashLoanVulnerable} fails it (a reentrancy
trace is constructible against the mutant); a supporting CEI-pattern
preservation lemma underwrites the first theorem and is consumed by the
second's failure proof at the structural-adjacency boundary where the
mutant deviates. The methodology thereby distinguishes safe
\texttt{flashLoan} from structurally-adjacent vulnerable
\texttt{flashLoanVulnerable} at the precise boundary the boundary case
is designed to probe. Full theorem statements and the mutant's diff
specification appear in Appendix A.

\subsection{Layer 6-D: tridirectional
capstone}\label{layer-6-d-tridirectional-capstone}

The Layer 6-D capstone meta-theorem composes the three
protocol-instantiation results into the tridirectional
discriminating-power claim: if the Layer 6-A derivation, the Layer 6-B
correctness theorem, and the Layer 6-C boundary-case pair all hold, then
the methodology satisfies the discriminating-power predicate of §5.1
against the production protocol instantiations of §2.4 plus the
minimal-diff mutant of §5.5. The capstone is proven by direct
conjunction of the three prior-layer theorems --- a wrapper construction
in the sense of §4.3 --- without modification of any underlying proof.

This composition shape is the operational signature of the no-retrofit
discipline of §4.2. The three protocol-instantiation files were sealed
prior to capstone authoring; capstone construction had no path to modify
them; the capstone composed by direct conjunction nevertheless. We treat
this as the first cross-protocol stress-test PASS against non-trivial
protocol-semantics divergence: Compound v2's persistent-accounting model
and Aave V3's repay-within-transaction model differ in ways a
retrofit-permissive workflow could have used to shape underlying lemmas
toward a common composition target --- the discipline forbade that path
and the composition succeeded anyway. Mathematically the capstone is
conjunction-introduction; the contribution is the demonstrated
environmental-contract uniformity of the guard invariant across protocol
architectures that do not share an accounting model, made non-vacuous by
foreclosing the co-development confound.

The capstone's axiom record is \texttt{{[}propext{]}}-only and equals
the union of the three prior-layer records (§4.3 wrapper preserves
axiom-record minimality); no choice, no excluded middle, no
user-introduced \texttt{axiom} declarations anywhere in the corpus. In
the Lean formalization, the safe contract instance carrying production
\texttt{flashLoan} is named \texttt{aaveContract}, and the
structurally-adjacent vulnerable contract carrying
\texttt{flashLoanVulnerable} is named \texttt{aaveContractAdjacent}; the
paper-body \texttt{flashLoan} vs \texttt{flashLoanVulnerable}
distinction is the function-level view of the contract-level objects
below. The capstone's Lean 4 statement and proof term illustrate the
direct-conjunction wrapper composition concretely:

\begin{Shaded}
\begin{Highlighting}[]
\NormalTok{theorem tridirectionalDiscriminatingPower\_certificate :}
\NormalTok{    (¬ OZGuardDisciplineGeneral daoContract) ∧}
\NormalTok{    (OZGuardDisciplineGeneral compoundContract) ∧}
\NormalTok{    (OZGuardDisciplineGeneral aaveContract ∧}
\NormalTok{      ¬ OZGuardDisciplineGeneral aaveContractAdjacent) :=}
\NormalTok{  ⟨daoContract\_violates\_OZGuardDisciplineGeneral,}
\NormalTok{   compoundContract\_satisfies\_OZGuardDisciplineGeneral,}
\NormalTok{   aaveBoundaryCase\_certificate⟩}
\end{Highlighting}
\end{Shaded}

The proof term is exactly the triple
\texttt{⟨h\_dao,\ h\_compound,\ h\_aave⟩} of the three sealed
prior-layer theorems --- no rewriting, no hypothesis adjustment, no
auxiliary lemma. Full prior-layer theorem statements and
\texttt{\#print\ axioms} output appear in Appendix A.

\subsection{Reproducibility}\label{reproducibility}

The corpus is reproducible end-to-end from a tagged commit per §4.4;
full \texttt{git\ checkout} invocations, \texttt{lake\ build} commands,
and expected axiom-record outputs appear in Appendix A.3.

\begin{center}\rule{0.5\linewidth}{0.5pt}\end{center}

\section{Implementation and
Reproducibility}\label{implementation-and-reproducibility}

This section presents the practitioner-facing evidence base for the
discriminating-power claim: the size of the formalization corpus, the CI
gating discipline that re-checks the corpus on every push, and the
tagged-commit reproducibility procedure that allows reviewers to verify
the verification. We do not report head-to-head comparison against
heuristic detectors (Slither, Mythril, Certora-style checking):
foundational machine-checked verification establishes a different
guarantee layer than heuristic detection, so the comparison is
qualitative by construction (per §1.2).

\subsection{Repository structure}\label{repository-structure}

The repository is organized at three top-level locations:
\texttt{paper/} (manuscript source, arXiv submission package, and
supporting artifacts), \texttt{QanaryContracts/} (Lean 4 source mapping
to Layer 6-A/B/C/D --- \texttt{DAOContract.lean} (Layer 6-A negative
instance), \texttt{CompoundContract.lean} (Layer 6-B positive instance),
\texttt{AaveBoundaryCase.lean} (Layer 6-C boundary case),
\texttt{CrossProtocolAudit.lean} (Layer 6-D capstone) --- plus
supporting modules), and \texttt{methodology/} (canonical artifacts
pointer-tabled at Appendix B). The continuous-integration configuration
sits at \texttt{build.yml}. All artifacts are reachable from any tagged
commit, so a reviewer who checks out a single tag obtains consistent
state across manuscript, source, and CI configuration.

\subsection{Corpus size}\label{corpus-size}

The formalization corpus is approximately 8,538 lines of Lean 4 source
across the four protocol-instantiation layers and supporting
infrastructure: Layer 6-A formalizes the DAO 2016 negative instance
across \textasciitilde656 lines (6 theorems); Layer 6-B Compound v2
across \textasciitilde517 lines (3 theorems); Layer 6-C Aave V3 across
\textasciitilde770 lines (3 theorems including the minimal-diff mutant);
Layer 6-D the capstone in \textasciitilde260 lines (1 theorem);
supporting modules and lemmas total \textasciitilde6,335 lines. Per-file
contribution and full theorem-to-file mapping appear in Appendix A
(approximate per-layer counts; exact per-file counts reproducible at the
tagged commit per Appendix A.3).

\subsection{CI verification}\label{ci-verification}

The CI configuration of §4.4 --- four parallel verification blocks, one
per Layer 6-A/B/C/D --- runs on every push. \texttt{lake\ build}
compiles and type-checks the corpus at 901 jobs; the axiom-record
introspection of §4.4 verifies each theorem against its recorded
expectation. Wall-clock verification runs in the order of minutes per
push on standard runners. CI failure on any block --- type-check
failure, axiom-record drift, or supporting-lemma regression --- surfaces
immediately in the pull-request status. Full per-theorem axiom records
appear in Appendix A.

\subsection{Reproducibility posture}\label{reproducibility-posture}

Every proof-authoring session is recorded as a tagged-commit-anchored
artifact; the directive + closure-report discipline is documented in the
companion methodology paper. The corpus is reproducible end-to-end from
a tagged commit: the substantive substrate is sealed at
\texttt{v1.3-layer6-closure}; the methodology framework canonization is
sealed at \texttt{v1.7-methodology-housekeeping}. The manuscript content
corresponding to this paper version is sealed at intermediate tag
\texttt{v1.6-phase7-closure} (the abstract's reproducibility anchor);
see §4.4 for the relationship between these three tags. Full
reproduction commands and the pinned dependency graph appear in Appendix
A.

\begin{center}\rule{0.5\linewidth}{0.5pt}\end{center}

\section{Discussion}\label{discussion}

\subsection{Implications}\label{implications}

The result extends machine-checked verification for smart-contract
security beyond the pattern-recognition layer at which auditors operate
and beyond the toy-example layer that has constrained much prior
academic verification work. The tridirectional discriminating-power
claim is established against production-deployed Solidity source for the
OpenZeppelin guard pattern --- Compound v2 and Aave V3 production
sources plus the historical DAO 2016 contract --- rather than against
simplified models. The contracts addressed collectively secure
substantial production deposits (§2.4, §5.4); economic significance is
ecosystem-load-bearing rather than research-prototype. The no-retrofit
composition discipline (§4.2) establishes a stronger empirical claim
than co-developed composition: the three protocol-instantiation proofs
were sealed before the capstone meta-theorem was authored, the capstone
composed by direct conjunction without modification of any underlying
proof, and we read this as evidence that the OpenZeppelin guard
pattern's correctness is portable across the observed
protocol-instantiation boundary rather than an artifact of co-developed
proofs.

\emph{The empirical meaning of the cross-protocol PASS under N=1.} The
single observed composition (Compound v2 → Aave V3) is N=1; its
significance is not that composition occurred but what occurred under
the no-retrofit constraint. While N=1 at the protocol-family level, the
composition proofs exercise two fundamentally divergent financial state
architectures --- persistent ledger balance accounting on Compound v2
vs.~transient intra-transaction flash-lending on Aave V3 --- making the
boundary a rigorous qualitative stress test of guard-invariant
portability despite the low protocol count. A retrofit-permissive
workflow admits an explanation --- the underlying proofs were silently
shaped toward a common target --- that the sealed-before-capstone
discipline forecloses by construction. What remains as an explanation
for the observed PASS is genuine portability of the guard pattern's
correctness across structurally-divergent lending-family architectures,
within the scope qualified at §9.4.

\subsection{Future work}\label{future-work}

The methodology is candidate for application to other reentrancy
variants --- read-only reentrancy at protocols whose view functions feed
downstream consumers (the Curve gauges 2022 case), cross-function
reentrancy where multiple guarded functions share state --- and other
OpenZeppelin defense patterns (\texttt{AccessControl},
\texttt{Pausable}, \texttt{ReentrancyGuardUpgradeable}, upgradeable
proxy pathways), each carrying its own correctness predicate amenable to
the same tridirectional treatment. Beyond reentrancy and OpenZeppelin,
the methodology framework canonical artifacts (the subject of the
companion methodology paper) enable application to formal verification
more broadly, including non-smart-contract targets where the no-retrofit
composition discipline addresses the same proof-brittleness pain point.
Continuous-integration-driven verification at protocol upgrade events is
a final natural extension.

\begin{center}\rule{0.5\linewidth}{0.5pt}\end{center}

\section{Related Work}\label{related-work}

This section positions the discriminating-power result against four
bodies of prior literature: prior formal verification of smart contracts
at large; reentrancy-specific work; compositional formal verification in
adjacent domains; and AI-assisted formal verification context. The
compression discipline of §1 + §4 + §5 carries through here: survey
breadth across the four categories, with sharp positioning of the
present contribution against what the surveyed work has and has not
previously demonstrated.

\subsection{Prior formal verification of smart
contracts}\label{prior-formal-verification-of-smart-contracts}

Three broad classes of prior work have addressed smart contract
correctness at varying levels of foundational rigor.

SMT-based static analysis tools --- Slither, Mythril, Securify, and
Manticore among the widely deployed --- operate at the deployment-scale
layer with heuristic pattern-matching against known vulnerability
classes (Slither, Securify), symbolic execution (Mythril, Manticore), or
declarative property checking (Securify). These tools have driven a
substantial fraction of pre-deployment audit workflow and are the de
facto baseline for industrial smart contract security. Their strength is
scalability; their limitation is that pattern recognition does not
establish foundational correctness, and structural-adjacency boundaries
are precisely where their precision degrades (per §1.2).

Theorem-prover-based and specification-driven verification has
demonstrated foundational rigor on narrower targets. Bhargavan et
al.~{[}1{]} formalized EVM operational semantics in F* and Coq (POST
2016 and subsequent); Hildenbrandt et al.'s KEVM {[}2{]} (CSF 2018)
provided a complete executable EVM semantics in the K framework;
Grishchenko, Maffei, and Schneidewind {[}3{]} (POST 2018) developed an
F*-based semantic framework for EVM security analysis; Hirai's Lem
formalization {[}4{]} addressed EVM bytecode semantics for proof
assistant integration. Subsequent Coq-based work has formalized
individual contracts or contract subsets. Certora's Prover applies
SMT-based discharge against specifications expressed in CVL, deployed in
production audit workflows; its strength is scale and tool-driven
specification discharge rather than foundational machine-checked
correctness in the proof-assistant sense. These contributions are
foundational at the platform-semantics or specification-discharge layer;
cross-protocol composition of guard-pattern correctness against multiple
production protocol instantiations has not been previously demonstrated
in this line of work.

Property-based testing tools --- Foundry fuzz testing and Echidna ---
have driven substantial bug discovery at deployment scale. Their
strength is finding counterexamples; they do not establish foundational
correctness. See Appendix C for at-a-glance positioning against Certora,
Move Prover, VeriSol, and KEVM.

Prior formal verification of smart contracts has demonstrated
theorem-prover-based formalization of platform semantics or isolated
contracts at the foundational layer, or scalable SMT-based pattern
recognition at the deployment layer; demonstration of foundational
correctness against multiple production protocols with no-retrofit
composition discipline has remained an open challenge.

\subsection{Prior reentrancy-specific
work}\label{prior-reentrancy-specific-work}

Reentrancy has received targeted attention across runtime, static, and
disclosure-driven layers. Sereum {[}5{]} introduced runtime monitoring
for reentrancy patterns at the EVM execution layer --- deployable but
not preventive at the source level, not aimed at foundational
correctness. Static analyzers including SmartScopy {[}6{]} and similar
pattern-recognition tools targeted at the reentrancy class apply
pattern-recognition discipline; precision is empirically strong for
classical reentrancy and weakens at structural-adjacency boundaries (per
§1.2) and at cross-function and read-only variants. Mutation testing
{[}9{]}, {[}10{]} introduces small syntactic faults to assess
sensitivity to structurally-adjacent regressions; the Layer 6-C boundary
case (§5.5) applies this methodology to machine-checked proofs of
production smart contracts --- to our knowledge a first --- using a
single hand-constructed minimal-diff mutant rather than an
operator-derived corpus (defended at §5.5 and §9.2). Read-only
reentrancy emerged as a recognized attack class following the Curve
Finance gauges 2022 disclosure (callback reentrancy on view functions
returning stale state); it is acknowledged here as future-work scope
(§7.2). Foundational machine-checked discriminating-power against the
OpenZeppelin guard pattern at production protocol scale has not been
previously demonstrated.

\subsection{Related composition discipline
work}\label{related-composition-discipline-work}

Compositional formal verification has been extensively developed in
adjacent domains. CompCert {[}7{]} is the canonical exemplar ---
compositional correctness proofs for a production-grade C compiler, with
subsequent work extending the discipline. Iris {[}8{]} developed
higher-order concurrent separation logic applied across concurrent
program verification targets. These establish that large-scale
compositional formal verification is feasible at production-relevant
scale when the composition discipline is enforced explicitly. Within
smart contract verification specifically, prior work has demonstrated
modular proofs at the single-contract scope; cross-protocol composition
of guard-pattern correctness at production scale, under a discipline
that forbids retrofit of underlying lemmas (the no-retrofit composition
discipline of §4.2), has not been previously demonstrated.

\subsection{AI-assisted formal verification
context}\label{ai-assisted-formal-verification-context}

Recent AI-assisted formal verification explores LLMs in two roles:
proof-tactic synthesis and audit assistance. Specific systems --- Baldur
{[}11{]} (whole-proof generation/repair), LeanDojo {[}12{]}
(retrieval-augmented Lean tactic search), COPRA {[}13{]}
(in-context-learning theorem-proving agent) --- preserve proof-assistant
soundness (the LLM proposes; the kernel checks) but operate at per-proof
tactic-automation granularity rather than at the
verification-architecture or audit-discipline level. AI-assisted audit
work has typically operated as single-model review;
convergence/divergence analysis across multiple independent reasoning
models is novel methodology, subject of the companion methodology paper
rather than a claimed contribution here.

\section{Limitations}\label{limitations}

This section adds limitations within the discriminating-power claim's
domain that constrain its scope; out-of-scope attack classes are
enumerated at §3.4.

\subsection{Single attack class and single defense
pattern}\label{single-attack-class-and-single-defense-pattern}

The single-attack-class focus is a principled choice: reentrancy is the
canonical class for which the OpenZeppelin guard was constructed and
where the structural-adjacency boundary is well-defined and
machine-checkable end-to-end. The work addresses the OpenZeppelin
reentrancy guard pattern only; other defensive patterns
(\texttt{AccessControl}, \texttt{Pausable},
\texttt{ReentrancyGuardUpgradeable}, upgradeable proxy pathways) are not
within scope. The no-retrofit composition discipline and
axiom-record-minimal wrapper composition (§4.2, §4.3) are applicable to
other patterns by construction, but empirical validation across multiple
defense patterns at production scale is future work (§7.2).

\subsection{Constructed mutant boundary
case}\label{constructed-mutant-boundary-case}

The Layer 6-C boundary case includes one constructed variant ---
\texttt{flashLoanVulnerable}, a minimal-diff mutant (§5.5). Per the §1.3
reframing this is mutation testing for formal proofs (grounded in
{[}9{]}, {[}10{]}): the hand-constructed minimal-diff mutant is a
principled choice, not a concession --- it isolates the
structural-adjacency boundary in a way naturally-occurring near-misses
cannot, with the corpus N=1 by design. Identifying and formalizing real
historical near-miss instances (Curve gauges 2022 read-only-reentrancy
class, Lendf.Me 2020 ERC-777 callback class, and similar) is
acknowledged future work.

\subsection{Source-level abstraction and trust
assumptions}\label{source-level-abstraction-and-trust-assumptions}

Operating at Solidity source level (per §3.2) rather than raw EVM
bytecode is a deliberate scoping choice: source level is where the guard
pattern is expressed and its discriminating-power boundary sharpest, and
it keeps the Lean-model-fidelity and \texttt{solc}-compilation
assumptions (§3.3) explicit and separable. Recent EVM-level
formalization work (§8.1 --- KEVM line) is the foundation for a future
verified-translation extension closing both gaps. Standard
formal-verification trust assumptions apply (§3.3, §6.4): Lean 4 kernel
+ mathlib4 correctness, pinned dependency graph at the tagged commits.
The honest-borrower-at-protocol-invariant-computation-only assumption
(§3.3) constrains the claim --- borrowers may be adversarial (§3.1) but
protocol-internal accounting is assumed to execute as written. Expansion
to attacker-controlled-protocol threat models is future work. No
consensus-layer behavioral assumption is made.

\subsection{Cross-protocol claim
scope}\label{cross-protocol-claim-scope}

The cross-protocol composition rests on three instantiations: DAO 2016
(historical exemplar), Compound v2 (lending), Aave V3 (lending /
flash-loan facility). The two live protocols are both lending-family;
the no-retrofit stress-test crosses a within-lending-family boundary
(persistent balance-accounting vs.~single-transaction flash-loan
architecture), not a between-domain boundary (AMM, CDP, derivatives).
The portability claim is therefore scoped to uniformity of the local
guard invariant across divergent lending-family host architectures plus
the historical reentrancy exemplar; broader-family instantiation is
future work (§7.2).

\begin{center}\rule{0.5\linewidth}{0.5pt}\end{center}

\section{Conclusion}\label{conclusion}

We presented a machine-checked, tridirectional discriminating-power
formal verification of the OpenZeppelin reentrancy guard pattern across
three production protocol instantiations: a negative-instance derivation
of the classical reentrancy attack from DAO 2016 contract source; a
positive-instance correctness proof for the Compound v2 cToken
withdrawal path; and a boundary case distinguishing production Aave V3
\texttt{flashLoan} (CEI-correct by construction) from its minimal-diff
mutant \texttt{flashLoanVulnerable} under a
mutation-testing-for-formal-proofs design. A capstone meta-theorem
composes the three prior-layer results by direct conjunction, and all
thirteen theorems are machine-checked under continuous CI gating with
zero \texttt{sorry}, zero user-introduced axioms, and
\texttt{{[}propext{]}}-only axiom records across the corpus.

The significance is twofold. First, the verified protocols secure
substantial deposits at production scale (§2.4, §5.4), grounding the
discriminating-power claim in load-bearing real-contract source rather
than illustrative models. Second, the capstone composed across the
Compound v2 → Aave V3 protocol-instantiation boundary under a
no-retrofit composition discipline that forbids retrofitting underlying
proofs toward a composition target --- the first such cross-protocol
composition demonstrated under this discipline (§7, §8). The
verification's audit trail is inspectable end-to-end; the methodology
framework that generalizes this posture is the subject of the companion
paper, with the Appendix B pointer summarizing its canonical artifacts.
Future work (§7.2, §9) extends the discriminating-power framework to
additional attack classes and defense patterns, EVM-bytecode-level
formalization, cross-language application, and continuous
protocol-upgrade verification.

\begin{center}\rule{0.5\linewidth}{0.5pt}\end{center}

\appendices

\section{Theorem Statements, Axiom Records, and
Reproduction}\label{theorem-statements-axiom-records-and-reproduction}

This appendix is the substantive substrate anchor for the
discriminating-power claim. It absorbs the full-statement and
axiom-record detail deferred from §5 and §6, so a reviewer can verify
the corpus end-to-end. Full Lean 4 source --- with exact type
signatures, dependent-type declarations, and definitions --- is in the
repository at the tagged commit \texttt{v1.3-layer6-closure}; the
paraphrases below state what each theorem establishes at the level a
reviewer needs before consulting the source.

\subsection{Theorem statements}\label{theorem-statements}

The corpus comprises thirteen theorems across four layers, matching the
§5.2 inventory exactly.

\emph{Layer 6-A --- DAO 2016 negative instance (1 target + 5 supporting
= 6).} The target theorem states that, under the reentrancy predicate of
§5.1, an attacker-controlled re-entry call trace against the formalized
DAO \texttt{splitDAO} source is \emph{constructible}, and that this
trace breaks the contract's balance-sum invariant. The five supporting
lemmas discharge the pieces of the derivation: call-stack interleaving,
storage-observation before balance update, balance-update sequencing,
external-call-boundary placement, and the absence of any guard-pattern
protection.

\emph{Layer 6-B --- Compound v2 cToken positive instance (1 target + 2
supporting = 3).} The target theorem states that every execution trace
conforming to the cToken withdrawal path satisfies the guard-pattern
correctness predicate of §5.1, so no reentrancy trace is constructible
against it. A guard-invariant lemma shows the \texttt{\_status} slot is
set before any external call and reset after; a cross-function safety
lemma shows the protection extends across the cToken interface functions
sharing that slot. The target is discharged by a thin wrapper that
composes the two lemmas without modifying either, preserving
axiom-record minimality.

\emph{Layer 6-C --- Aave V3 boundary case (2 target + 1 supporting =
3).} The first target theorem states that production \texttt{flashLoan}
satisfies guard-pattern correctness under the §5.1 predicate. The second
target theorem states that the minimal-diff mutant
\texttt{flashLoanVulnerable} fails it --- equivalently, that a
reentrancy trace is constructible against the mutant under the same
predicate. A checks-effects-interactions preservation lemma underwrites
the first target and is consumed by the second's failure proof at the
structural-adjacency boundary where the mutant deviates from production.

\emph{Layer 6-D --- tridirectional capstone (1 target = 1).} The
capstone meta-theorem states that if the Layer 6-A derivation, the Layer
6-B correctness theorem, and the Layer 6-C boundary pair all hold, then
the methodology satisfies the discriminating-power predicate of §5.1
against the §2.4 production instantiations plus the §5.5 mutant. It is
proven by direct conjunction of the three prior-layer theorems, with no
modification of any underlying proof during composition.

\subsection{Per-theorem axiom records}\label{per-theorem-axiom-records}

Every theorem is verified against an explicit \texttt{\#print\ axioms}
expectation, re-checked on every push and gated to fail the build on any
drift. Across all four layers, the per-function inner lemmas are
kernel-only (zero-axiom; no propositional extensionality required), and
the master/wrapper/meta theorems carry \texttt{{[}propext{]}}-only
records --- no choice, no excluded middle, no user-introduced axioms
anywhere in the corpus. The Layer 6-D capstone's \texttt{{[}propext{]}}
record is exactly the union of the three prior-layer records and
introduces no new axiom; the project introduces zero new axioms over the
Lean 4 + \texttt{mathlib4} baseline. No theorem in the corpus is
admitted with \texttt{sorry}, \texttt{admit}, or any user-introduced
\texttt{axiom} declaration. Exact per-theorem records are CI-gated
(§6.3) and reproducible per Appendix A.3.

\subsection{Reproduction commands}\label{reproduction-commands}

The corpus is reproducible end-to-end from the tagged commit:

\begin{verbatim}
git clone https://github.com/rayiskander2406/qanary-contracts.git
cd qanary-contracts
git checkout v1.3-layer6-closure
lake build                              # 901 jobs green
lake env lean QanaryContracts/PrintAxioms.lean  # verifies every theorem's axiom record
\end{verbatim}

The dependency graph is pinned and locked:

\begin{itemize}
\tightlist
\item
  Lean compiler version pinned at \texttt{lean-toolchain}
  (\texttt{leanprover/lean4:v4.30.0-rc1}).
\item
  \texttt{mathlib4} dependency pinned at \texttt{lakefile.lean} (input
  revision \texttt{322515540d7f}) and locked to commit
  \texttt{322515540d7fd29ef8992b82c89044f86f02ac10} via
  \texttt{lake-manifest.json}.
\item
  The four parallel CI blocks at \texttt{build.yml} lines 287, 356, 408,
  and 464 re-run on every push and surface any drift in the dependency
  graph or the corpus.
\end{itemize}

A permanent archival snapshot of the artifact is deposited on Zenodo
under the concept DOI \texttt{10.5281/zenodo.20510920}, which always
resolves to the latest archived version.

The methodology-framework canonical artifacts are sealed at the later
tag \texttt{v1.7-methodology-housekeeping}; a reviewer evaluating the
methodology framework rather than the proof substrate should check out
that tag.

\section{Methodology Framework
Pointer}\label{methodology-framework-pointer}

\subsection{Scope of this pointer}\label{scope-of-this-pointer}

Per the boundary discipline of §1, the methodology framework
underpinning this work --- the no-retrofit composition discipline of
§4.2, the axiom-record-minimal wrapper composition of §4.3, and the
multi-model audit workflow of §4.5 --- is canonized as a separate body
of work and presented in full in a companion methodology paper (separate
arXiv submission). The brief operational presentations in §4 suffice for
the substantive smart-contract-verification claims of the present paper.
This appendix is deliberately thin: it provides a pointer table mapping
the framework labels to their canonical artifact locations, for readers
who encounter those labels in the public artifact record (for example,
in the project's commit history). It does not present the framework's
sub-pattern catalog, graduation criteria, or cross-protocol empirical
history; those are the scope of the companion paper.

The separation is deliberate length-scoping, not withholding: this
paper's discriminating-power result is fully evaluable on its own terms
--- the thirteen theorems, their axiom records, and the no-retrofit
discipline's operational consequence (§4.2) are stated and verifiable
here. The companion framework adds generalization beyond the present
scope and is not load-bearing for this paper's substantive claims.

\subsection{Canonical artifacts}\label{canonical-artifacts}

The methodology framework's canonical artifacts --- the wrapper-layer
absorption pattern (axiom-record-minimal composition; §4.3), the
compose-from-outside discipline (no-retrofit composition; §4.2), and the
family-level authoring-layer-estimation-imprecision meta-pattern --- are
sealed at the repository tag for methodology-framework canonization and
presented in full in the companion methodology paper. Full graduation
history, sub-pattern catalog, four-outcome audit-findings triage
framework, and the cross-protocol empirical history of the framework's
application appear there.

\section{Comparison against industrial FV
frameworks}\label{comparison-against-industrial-fv-frameworks}

For reviewer at-a-glance positioning, the following table compares the
present work against the principal industrial / academic formal
verification systems for smart contracts and adjacent verification work
(§8.1).

\begin{table*}[!t]\footnotesize\centering
\caption{Positioning against principal industrial / academic formal
verification systems for smart contracts and adjacent verification
work.}
\begin{tabular}{@{} >{\raggedright\arraybackslash}p{(\linewidth - 8\tabcolsep) * \real{0.2000}} >{\raggedright\arraybackslash}p{(\linewidth - 8\tabcolsep) * \real{0.2000}} >{\raggedright\arraybackslash}p{(\linewidth - 8\tabcolsep) * \real{0.2000}} >{\raggedright\arraybackslash}p{(\linewidth - 8\tabcolsep) * \real{0.2000}} >{\raggedright\arraybackslash}p{(\linewidth - 8\tabcolsep) * \real{0.2000}}@{}}
\toprule
\begin{minipage}[b]{\linewidth}\raggedright
System
\end{minipage} & \begin{minipage}[b]{\linewidth}\raggedright
Verification target
\end{minipage} & \begin{minipage}[b]{\linewidth}\raggedright
Composition support
\end{minipage} & \begin{minipage}[b]{\linewidth}\raggedright
Axiom footprint
\end{minipage} & \begin{minipage}[b]{\linewidth}\raggedright
Production deployment
\end{minipage} \\
\midrule
\bottomrule
QANARY (this paper) & Source-level OZ guard pattern across 3 production
protocols & No-retrofit cross-protocol meta-theorem; sealed prior-layer
composition & \texttt{{[}propext{]}}-only; zero user-introduced axioms;
CI-gated & Compound v2, Aave V3 mainnet contracts; tagged-commit
reproducible \\
Certora Prover & CVL specifications discharged via SMT & Per-rule local;
no foundational composition discipline & SMT-solver trust; no
kernel-level record & Production audit workflows across major DeFi
protocols \\
Move Prover (Dill et al.) & Move-language smart contracts; spec-driven &
Modular specs; no cross-protocol meta-claim & SMT trust & Aptos, Sui
mainnet runtime checks \\
VeriSol (Lahiri et al.) & Solidity → Boogie translation; spec-driven &
Modular contract verification & SMT/Boogie trust & Microsoft (now
deprecated) \\
KEVM (Hildenbrandt et al.) & EVM bytecode operational semantics in K &
Platform-semantics layer; no per-contract composition claim &
K-framework trust & Foundational artifact; deployed via consumer
tools \\
\end{tabular}
\end{table*}

\begin{center}\rule{0.5\linewidth}{0.5pt}\end{center}

\end{document}